\documentclass[aip,apl,reprint,twocolumn,showpacs,showkeys]{revtex4-1}
\usepackage{graphicx}
\usepackage{amsmath}

\begin{document}
\title{Solution of Maxwell's Equations}
\author{Seoktae \surname{Lee}}
\email{stlee@semyung.ac.kr}
\affiliation{Electronic Engineering Department, Semyung University, Jecheon
390-711}
\date[]{Received August 2010}


% latex definition ................................................
\def\a{\alpha}               \def\b{\beta}
\def\d{\delta}               \def\D{\Delta}
\def\e{\epsilon}
\def\f{\phi}                 \def\F{\Phi}
\def\g{\gamma}            \def\G{\Gamma}
\def\j{\psi}                  \def\J{\Psi}
\def\l{\lambda}             \def\L{\Lambda}
\def\m{\mu}                 \def\n{\nu}
\def\o{\omega}               \def\O{\Omega}
\def\p{\pi}                  \def\P{\Pi}
\def\r{\rho}
\def\s{\sigma}              \def\S{\Sigma}
\def\t{\tau}
\def\th{\theta}             \def\Th{\Theta}

\def\etal{{\it et al.}}

\def\sl #1{\not\!{#1}}

\def\abs#1{\left| #1\right|}
\def\Bar{\overline}

\def\fr{\frac}
\def\cd{\cdot}
\def\ba{\begin{array}}          \def\ea{\end{array}}
\def\bz{\begin{equation}}       \def\ez{\end{equation}}
\def\by{\begin{eqnarray}}       \def\ey{\end{eqnarray}}
\def\pa{\partial}               \def\na{\nabla}
\def\tb{\textbf}
\def\nn{\nonumber}              \def\ni{\noindent}

%Roman Numeral
\makeatletter
\newcommand{\rmnum}[1]{\romannumeral #1}
\newcommand{\Rmnum}[1]{\expandafter \@slowromancap\romannumeral #1@}
\makeatother

%.........................................................................
\begin{abstract}
The Maxwell's equations are solved when it has an inhomogeneous terms as a
source. The solution is very general in a sense that it handles arbitrary
current source and anisotropic media. The calculation is carried out in the k-domain after Fourier transform, and its results are confirmed with 
the propagators of the new coupling-free wave equations derived from the Maxwell equations for anisotropic environments.
.
\end{abstract}
\maketitle

The set of Maxwell's equations is a classical representation of electromagnetic
field theory. It is still a center of technological applications as well as
theoretical research in many areas of physical science.  The physical prediction of wave propagation is made through the fields calculation and its basis is the Maxwell's and wave equations. The task obtaining the solution of the Maxwell's equations is important as much as understanding the the equations themselves.\\

The Maxwell equations are the first order, linear, simultaneous partial differential equations(PDE) for vector fields {\tb E} and {\tb H}, but their solution is not easily obtained. The main difficulty lies in the fact that the unknown variables {\tb E} and {\tb H} are coupled to each other in the equations. The most frequently cited method of solving the Maxwell's equations is to take a curl operator on both sides of Maxwell's 3rd(Faraday's law) and 4th(extended Ampere's law) equations. This operation is specially powerful for an isotropic medium, since it gives a decoupled version of wave equations with source terms. These wave equations are not only the equations for waves, but also solvable forms of the Maxwell's equations. The wave equations of the isotropic medium have been solved conventionally by Green's function technique.\\

The solution of the wave equations has a long history. The practical advancements have been achieved in 1950s by Schwinger and Marcuvitz, who solved first the inhomogeneous wave equations  using Green's functions\cite{sch}. Soon after, Many other scholars, for example, F. V. Bunkin\cite{Bunkin}, C. T. Tai\cite{Tai}, and L. B. Felsen\cite{Felsen} developed methods using the so-called dyadic Green's functions, which have been used to this time. \\

The tensor(or dyadic) Green's functions have been widely studied specially last decade, as the number of electronic equipment operating in an anisotropic environment increased. However, the general solution of Maxwell's equations or wave equations in an anisotropic medium have not been known because of coupling problem between {\tb E} and {\tb H}. The coupling problem of the equations for the anisotropic medium is more complicate than the isotropic case, since the permittivity $\e$ and permeability $\m$ are matrices. The general wave equations for an anisotropic medium become
\by \na \times \na \times {\bf E} &=&  - \frac{\pa
}{{\pa t}}(\na  \times \mu {\bf H}), \\
\na \times \na \times {\bf H} &=& \na \times \bf J +
\frac{\pa}{\pa t} (\na \times \e \bf E). \ey
It is noted that the first term of Eq.(1) was not expanded to $\na(\na\cdot {\bf E}) - \na ^2 {\bf E}$, as the  $\na\cd{\bf E}$ cannot be written as $\e^{-1} \r$ when applying the Gauss's law. The correct form of the first Maxwell's equation is $\na  \cdot \e {\bf E} = \rho$, instead of $\na  \cdot {\bf E} = \e^{ - 1} \rho$, since $\e$ is a matrix. The $\e$ cannot move out from the inside of divergence. It is also important to point out that  the $\e$ and $\m$ in the curl operator of Eq.(1) and (2) cannot come to the front of the curl operator, as a constant number does. Thus, the equations can be solved only within a restricted range. In an arbitrary general case, the equations are practically regarded as insolvable without using  numerical simulation as far as we follow usual conventional ways.\\

This paper attempts general solutions of the fields {\tb E} and {\tb H} in 2 different ways. The first method is to solve the fields directly from the Maxwell's equations, and the other one is to apply the Green's function technics on the new wave equations, which are derived in a different way to fit into an anisotropic environment. The two solutions, which are obtained respectively in different ways, turn out to be same.\\

It is useful to write down the Maxwell's 3rd and 4th equations in an unifying description method as following, 
\by &&\na \times
\left( {\begin{array}{*{20}c}
   {\bf E} \\
   {\bf H} \\
\end{array}} \right) + {\Big\{}
\fr{\pa}{\pa t}
\left( {\begin{array}{*{20}c}
   \m {\bf H} \\
   -\e {\bf E} \\
\end{array}} \right) \;  {\Big\}}\nn\\
=\; && {\Big\{}\na \times
 + 
\fr{\pa}{\pa t}
\left( {\begin{array}{*{20}c}
   0 & \m \\
   -\e & 0 \\
\end{array}} \right)
\;  {\Big\}} \cd
\left( {\begin{array}{*{20}c}
   {\bf E} \\
   {\bf H} \\
\end{array}} \right) =\; 
\left( {\begin{array}{*{20}c}
   0 \\
   {\bf J} \\
\end{array}} \right). \ey\\

By using harmonic time dependency, the above equation becomes\\
\bz {\Big\{} \na \times  + i \o
\left( {\begin{array}{*{20}c}
    0 & \m \\
   -\e & 0 \\
\end{array}} \right)
{\Big\}} \cd
\left( {\begin{array}{*{20}c}
   {\bf E} \\
   {\bf H} \\
\end{array}} \right)\; =\;
\left( {\begin{array}{*{20}c}
   0 \\
   {\bf J} \\
\end{array}} \right). \ez \\
{\tb E}, {\tb H}, and {\tb J} in this equation are now the function of spatial coordinates. The equation is separated in 3 parts - operator, unknown {\tb E} and {\tb H}, and inhomogeneous term as a source. The operator part in the bracket acts on fields $\left( {\begin{array}{*{20}c} {\bf E} \\ {\bf H} \end{array}} \right)$, which constitutes a linear simultaneous PDE. The Fourier transformation has an advantage in transforming a linear differential equation into a corresponding algebraic equation. The curl operator is replaced by ($-i{\bf k} \times $), but this vector notation is not convenient in conjunction with matrix structure. If we use the following matrix $\tilde k$ representing wave vector $\bf k$ \cite{abdul}\\
\bz \tilde k = \left( {\begin{array}{*{20}c}
   0 & { - k_x } & {k_y }  \\
   {k_z } & 0 & { - k_x }  \\
   { - k_y } & {k_x } & 0  \\
\end{array}} \right), \ez

then ${\bf k} \times {\bf A} = \tilde k \cd {\bf A}$ for any arbitrary vector
${\bf A}$. Therefore the Eq.(4) becomes

\bz
\left( {\begin{array}{*{20}c}
   \tilde k & -\o \m \\
   \o \e & \tilde k \\
\end{array}} \right)
\left( {\begin{array}{*{20}c}
   {\bf E} \\
   {\bf H} \\
\end{array}} \right)\; =
\left( {\begin{array}{*{20}c}
   0 \\    i{\bf J} \\
\end{array}} \right). \ez
Here, $\tilde k$ is defined in rectangular coordinates. This implies that
$\tilde k$ is coordinates dependent. One can find it easily  in spherical or
cylindrical coordinates. The quantities ${\bf E},\; {\bf H},\; {\rm and} \; {\bf
J}$ in the equation are Fourier transformed functions of {\tb k}. The first
$2\times 2$ matrix is actually a coefficient matrix of the simultaneous
equation, but it should be taken care of the fact that each component of matrix is $3\times 3$ matrix. The 3rd and 4th Maxwell's equations become simultaneous equations. The solution is calculated by diagonalization of the coefficient matrix.
\bz
\left( {\begin{array}{*{20}c}
   \tilde k \cd \m^{-1} \tilde k + \o^2 \e & 0 \\
    0 & \tilde k \cd \e^{-1} \tilde k + \o^2 \m \\
\end{array}} \right)
\left( {\begin{array}{*{20}c}
   {\bf E} \\
   {\bf H} \\
\end{array}} \right)\; =
\left( {\begin{array}{*{20}c}
 i \o {\bf J} \\ i \tilde k \e^{-1} {\bf J} \\
\end{array}} \right), \ez
or equivalently
\by &&{\bf E}({\bf k}) =
      (\tilde k \cd \m^{-1} \tilde k + \o^2 \e)^{-1} \cd i\o {\bf J}({\bf k}) \\
    &&{\bf H}({\bf k}) = (\tilde k \cd \e^{-1} \tilde k + \o^2 \m)^{-1} \cd    
i\tilde k \e^{-1} {\bf J}({\bf k}) \ey
As {\tb E} and {\tb H} are functions of {\tb k}, the original spatial fields ${\bf E}({\bf r},{\bf r'})$ and ${\bf H}({\bf r},{\bf r'})$ are
obtained by an inverse Fourier transformation. The spatial points ${\bf r},\; {\bf r'}$ are each point of field and source. Now, the solutions are derived directly from the Maxwell's equations, not from the wave equations. Moreover, the application range of the solution is extended to an arbitrary anisotropic media.\\

As seen above, the simple algebraic manipulation of Maxwell's equations leads to
the solutions. If one takes look at the Eq.(7) carefully, the right term
$\left( {\begin{array}{*{20}c}
 i \o {\bf J} \\ i \tilde k \e^{-1} {\bf J} \\
\end{array}} \right)$
plays a role of source term, and the components of the first matrix in the left
side can be interpreted as Fourier-transformed Green's functions or 3
dimensional propagators of the fields. Usually, the anisotropic propagator of
the electromagnetic fields are available only in special cases, but the solution
(8) and (9) provides the general propagators as follows.
\by &&g_E ({\bf k}) = (\tilde k \cd \m^{-1} \tilde k + \o^2 \e)^{-1} \\
    &&g_H ({\bf k}) = (\tilde k \cd \e^{-1} \tilde k + \o^2 \m)^{-1} \ey
The fact that we find the general propagators implies we can calculate fields
only by integration. Thus, the solving procedure of the electromagnetic fields
can be automated.\\

If these solutions, directly obtained from the Maxwell's equations, are right,
the wave equations should also give the same results. However, $\e$ and $\m$ are replaced with matrices in the anisotropic medium, and the wave equations cannot be decoupled in general. The problem is essentially due to the fact neither $\e$ nor $\m$ can be extracted from the vector differential operators. Unlike an isotropic case, it is impossible to apply Maxwell's equations on $\na \times \e {\bf E }$ and $\na \times \m {\bf H}$. The terms $\e \cd \bf{E}$ and $\m \cd \bf{H}$ are merely a linear combination of the original $\bf{E}$ and $\bf{H}$, but they act as another unknown independent variables, when contained inside a vector differential operator. With Eq.(1) and (2), the decoupled version of wave equations is not possible. There are some examples of solvable models even in the anisotropic medium, when the anisotropic characteristic comes from either $\e$ or $\m$\cite{Lee}. However, the general problems, when anisotropy occurs due to both $\e$ and $\m$, can not avoid the coupling problem. Whether they are solvable or not depends on whether they become decoupled or not. The solution of the general problems doesn't seem to be feasible as far as the ordinary wave equations are used.\\

At this stage, the goal is obviously to find the decoupled version of wave
equations capable of solving the fields {\tb E} and {\tb H}. A better approach
is to go back to the original Maxwell's equations and derive new equations
instead of relying on the original wave equations. The magnetic field {\tb H} can be written from the 3rd Maxwell's equation as
\bz {\bf H} = \int - \m^{-1}\cd (\na \times {\bf E})\; dt.\ez
Inserting the above equations into the 4th Maxwell's equation and differentiating with respect to
time gives
\bz \na\times (\m^{-1} \cd (\na\times {\bf E})) + \; \e \frac{\pa^2 {\bf E}}{\pa
t^2} = -  \fr{\pa {\bf J}}{\pa t}\; .\ez
This equation is a new type of a wave equation for the field {\tb E}. It is
decoupled and doesn't have {\tb H}. A new wave equation for {\tb H} can be
obtained in a similar way. The result reads
\bz \na\times \e^{-1} (\na \times {\bf H}) + \m \;
\frac{\pa^2 {\bf H}}{\pa t^2} = \na\times (\e^{-1} {\bf J}). \ez
Now, these two equations Eq.(13) and (14) constitute new wave equations\cite{Lee} instead of the original wave equations for the isotropic case. These new equations are decoupled even in any arbitrary anisotropic environment, and consequently solvable. Moreover, they recover their original wave equation forms when there is no anisotropy. The next step is to check whether or not the propagators of the new and the original wave equations are identical to each other. The time dependency is again assumed to be harmonic. Then the propagators of Eq.(13) and (14) are obtained from the spatial Green function equations replacing source terms with the $\d$ functions.  The Green's function $G_E ({\bf r},{\bf r'})$ and $G_H ({\bf r},{\bf r'})$ are
\by &&\na\times (\m^{-1} \cd (\na\times {\bf G_E} )) - \o^2\e \; {\bf G_E} =
\d({\bf r} - {\bf r'}),\\
&& \na\times (\e^{-1}\cd (\na \times {\bf G_H})) - \o^2 \m \;
{\bf G_H} = \d({\bf r} - {\bf r'}). \ey
One of the advantage of Fourier transformation is in switching a linear
differential equation to an algebraic equation. The propagator is the Fourier
Transform of spatial Green's functions.
\by && G_E ({\bf r},{\bf r'})  = \int g_E ({\bf k}) e^{-i {\bf k}\cd ({\bf   
r}-{\bf r'})} d^3 {\bf r}\\
&& G_H ({\bf r},{\bf r'})  = \int  g_H ({\bf k}) e^{-i {\bf k}\cd ({\bf      
r}-{\bf r'})} d^3 {\bf r}\ey
The {\tb k}-space algebraic equations for propagators $g_E ({\bf k})$ and $g_H
({\bf k})$ are obtained by inserting above equations into Eq.(13) and (14).
\by &&g_E ({\bf k}) = -({\tilde k} \m^{-1} {\tilde k} + \o^2 \e )^{-1} \\
    &&g_H ({\bf k}) = -({\tilde k} \e^{-1} {\tilde k} + \o^2 \e )^{-1} \ey
Comparing with the results obtained in the previous paragraph, there exists (-1)
times difference overall in both $g_E$ and $g_H$. These differences come about
due to (-1) times differences in the source terms. The both results are exactly
same. \\

The two equation sets, Maxwell's equations and wave equations, which look
different in the {\tb r} space, turn out to be same in the {\tb k} space. The
fact that two equations have the same solutions has an mathematical implication.
The solution of Maxwell's equation has been obtained through the diagonalization
of the matrix, but the other solution of the wave equation has been treated with
the Green's function technique. The coincidence of two solutions indirectly
shows supports through the mutual verification of both solutions.\\

As for the justification or the proof of the solutions, a few things can be
mentioned. Firstly, the new wave equations become the original wave equations,
when there is no anisotropy. It implies that the new equations are the extension
of the original wave equation to the anisotropic environment. Secondly there are
a few number of researches, whose results are consistent with that of this
article. For example, Cottis, Vazouras and Spyrou\cite{Cottis} calculated the
Fourier-transformed dyadic Green's functions in an electric anisotropy, and
their result reads
\bz gE_{CVS}({\bf k}) = \frac{1}{{( (k^2{\bf I} - {\bf k} \otimes {\bf k})
- k_0^2 \mu _r \epsilon _r )}}, \; \ez
where $\e_r$, and $\m_r$ are the relative matrix coefficients as usual.
The result of the new equation for an identical case is given below according to
Eq.(19):
\bz gE_{NEW}({\bf k}) = \frac{-1}{{\tilde k}\; {\tilde k} +k_0^2 \; \e_r}\; .\ez
Given that there is no anisotropy in $\m$, the identity matrix is substituted in
place of $\m_r^{-1}$. The same result is obtained using the relationship $\tilde
k \cd \tilde k \; = \; {\bf k}\otimes {\bf k}-k^2 I$. In addition, calculation
of magnetic anisotropy and conditions of refractive index imposed on plain wave
also show identical results between two systems. This coincidence is natural
consequences, because they are based on the same mathematical ground.\\

In conclusion, we have shown the general solution of Maxwell's
equations with source terms in 2 different ways. The solution of the first
attempts  was calculated directly from the Maxwell's equations using matrix
diagonalization of operator part after Fourier transformation. The second method
is to solve the newly derived coupling free wave equations with the tensor Green's functions.  The two solutions, from matrix diagonalization and Green's functions, turn out to be identical, and are coincident with previous researches. These solutions are expected useful in many practical field calculations including inverse problems.

\end{document}